\documentstyle[12pt,fleqn]{article}
\input epsf
\def\singlespace {\smallskipamount=3.75pt plus1pt minus1pt
                  \medskipamount=7.5pt plus2pt minus2pt
                  \bigskipamount=15pt plus4pt minus4pt
                  \normalbaselineskip=15pt plus0pt minus0pt
                  \normallineskip=1pt
                  \normallineskiplimit=0pt
                  \jot=3.75pt
                  {\def\smallskip {\vskip\smallskipamount}}
                  {\def\medskip   {\vskip\medskipamount}}
                  {\def\bigskip   {\vskip\bigskipamount}}
                  {\setbox\strutbox=\hbox{\vrule
                    height10.5pt depth4.5pt width 0pt}}
                  \parskip 7.5pt
                  \normalbaselines}

\def\doublespace {\smallskipamount=7.5pt plus2pt minus2pt
                  \medskipamount=15pt plus4pt minus4pt
                  \bigskipamount=30pt plus8pt minus8pt
                  \normalbaselineskip=30pt plus0pt minus0pt
                  \normallineskip=2pt
                  \normallineskiplimit=0pt
                  \jot=7.5pt
                  {\def\smallskip {\vskip\smallskipamount}}
                  {\def\medskip   {\vskip\medskipamount}}
                  {\def\bigskip   {\vskip\bigskipamount}}
                  {\setbox\strutbox=\hbox{\vrule
                    height21.0pt depth9.0pt width 0pt}}
                  \parskip 15.0pt
                  \normalbaselines}
\def\be{\begin{equation}}
\def\ee{\end{equation}}
\def\bea{\begin{eqnarray}}
\def\eea{\end{eqnarray}}

\def\sect #1{\setcounter{equation}{0}}

\begin{document}
\singlespace
\doublespace

\centerline {\Large{Naked Singularity, Black-hole and the Mass
 Loss in a }}
\centerline {\Large{Spherically Symmetric Gravitational Collapse }}

\vspace{12pt}

\centerline{I H Dwivedi$^{*}$}
\centerline{21, Ballabhji Vihar, Dayalbagh, Agra, India}

\vspace{0.4in}
\vspace{2in}

\vspace{0.4in}

\noindent $*$\\
\noindent E-Mail Address:idwivedi@sancharnet.in;
dwivedi@tifr.res.in

\newpage

\vspace{1.3in}

\begin{abstract}
A distinguishable physical property between a naked singularity
and a black-hole, formed during a gravitational collapse has
important implications for both experimental and theoretical
relativity. We examine the energy radiated during the spherically
symmetric Gravitational collapse in this context within the
framework of general relativity. It is shown that total energy
radiated( Mass Loss) during the collapse ending in a naked
singularity scenario cannot be more than the case when a collapse
scenario ends in a black-hole. In cases of interest(for example
stars having same mass, size, and internal composition) considered
here the total energy released in the collapse ending in a black
hole can be considerably more than the case otherwise.

\end{abstract}

\newpage \section{Introduction}
The universe is full of supper high energy events. For example
quasars(and other active galactic nuclei) radiating $10^3$ times
more than the luminosity of a galaxy, supernova explosions
emitting huge amounts of energy within a few days to a year and
gamma ray bursts which radiate energy of the order of
$10^{51}$-$10^{54}$ ergs within a duration of few milliseconds to
few seconds. Despite numerous attempts, an ideal explanation is yet
to be found. It is, however, widely believed that probably
gravitational collapse is responsible for such events and that
during a dynamic collapse huge amount of energy would be released,
possibly due to the conversion of the rest mass of the star into
energy. A gravitational collapse
occurs when a star with sufficient mass has exhausted its nuclear
fuel. As this stellar collapse progresses further, gravitational
forces become very strong resulting in the development of closed
trapped surfaces. The development of this surface signals the
formation of a region of space time from where no causal particles
could escape. According to singularity theorems a space time
singularity must develop generically during a gravitational
collapse specially once a trapped surface develops. These
theorems, however do not say any thing about the nature of such
singularities and its physical properties.

With the advent of cosmic censorship conjecture (CCH)and
subsequent development of Black-hole physics, efforts have been
devoted over the last two decades on the study of the dynamics of
a gravitational collapse of physically reasonable matter in the
context of CCH. Conclusions from these are that singularities both
naked and covered would form generically during the collapse(see
\cite {D1,D2,D3} and the references therein). The implication of
these studies is that even if the collapse produces unbounded
curvature and density, trapped surface may not develop early to
prevent the exposure of the singular regions to an outside
observer.

Considerable work has appeared in the literature which attribute
these high energy events(for example gamma ray bursts) in the
cosmos to the formation and existence of a naked singularity
during the gravitational collapse ( See \cite{j1,j2} and further
references therein). The suggestion is that contrary to the black
hole scenario, in case of naked singularity, the high curvature
region of the space time( where the naked singularity is formed)
is open to the outside and therefore energy of this region some
how (either due to quantum particle production\cite {j2,j3} or
formation of a compact fireball \cite {j1})would lead to the high
energy emission from the vicinity of the naked singularity and
onward to the outside world. This is quite natural and one would
expect that since an extra high curvature region of the space time
with extreme conditions is available to radiate energy in the case
of a naked singularity, the energy released should be more than
the case of a black hole. Thus this could also possibly be a
observable signature of naked singularity.

Based on CCH Black-hole physics has developed significantly.
Similarly could the formation of a naked singularity also signal a
physical phenomena responsible for huge amounts of energy release.
An important aspect is  whether the occurrence of a singularity(
naked or otherwise) is just a mathematical problem due to the
mathematical structure of the field equations or do they have a
deeper physical meaning?. Investigation, therefore, of the two
scenarios namely naked singularity and black-hole from the
perspective of the total mass loss(energy radiated) during the
gravitational collapse is of theoretical interest in it's own
right in  general relativity. Moreover is there a distinguishable
difference between the amount of energy released during a collapse
process which is ending either as a black hole or a naked
singularity under similar conditions. The implication of such a
study within the framework of general relativity could not be
overemphasized.

We make such an attempt in this paper  and consider spherically
symmetric gravitational collapse from the point of view of
examining the energy radiated(Mass Loss) during the collapse
ending  in a singularity within the framework of general
relativity. We calculate the remnant mass of the collapsing star
which is shown to be related to the initial mass and the initial
pressure at the boundary of the star. It is then shown that the
total energy radiated in the model of a collapsing star considered
here can be same for both end states(naked singularity or black
hole) i.e. One can have, for equal amount of energy radiated
during the collapse, different sets of parameters at the onset of
collapse leading the star either to a naked singularity or a black
hole. Thus a naked singularity scenario in general would not
radiate more energy than a black hole scenario. An equally
important result is that for two stars having the same mass, size,
and the internal structure, the star ending in a black hole would
suffer a greater mass loss then its counterpart that is naked
singularity.

Here is the summary of what we intend to do and how we organize
rest of our paper. We would of course consider the collapse
scenario within the framework of classical general relativity. A
typical scenario involves a massive star on a continuous
gravitational collapse starting at some time $t=t_i$. In the next
section we therefore consider such a spherically symmetric matter
cloud representing type I matter field \cite{D4}. The radiating
Vaidya space time describes the space-time outside the moving
boundary of such a collapsing star. The conditions on the initial
density etc. are then  examined for the collapse to end either as
a black-hole or a globally naked singularity. Next in the section
on Mass Loss we discuss the mass function in Vaidya space-time and
it's relation to the interior parameters. The remnant mass of the
star is then calculated in terms of mass, density and pressure of
the star at the onset of collapse. In appendix A we discuss the
interior matter field in the context of field equations and energy
conditions. Appendix B is devoted to the star model and the
boundary conditions.

\newpage
\section{The Collapsing Star}

Despite the numerous exact solutions of the  field equations, very
few describe a physically reasonable collapsing matter cloud. In
fact in nearly all the studies of spherically symmetric collapse,
Tolman-Bondi-Lemitre metric(TBL) \cite{TBL} has been used
extensively and very well may be the only reasonable exact
solution available. While dust is a possible equation of state
describing late stages of collapse \cite{Pen}, it is an idealized
form of matter with vanishing pressure . Total mass enclosed in
the dust cloud is time independent(Exterior space-time is
Schwarzschild) and no energy is radiated during such a collapse.
We therefore consider a general matter field with non vanishing
pressures.
The metric describing a spherically symmetric space-time is
given by
\be
ds^2= -dt^2 +{R'^2\over 1+f}dr^2 + R^2d\Omega^2,
\ee
where
$d\Omega^2=d\theta^2+\sin^2\theta d\phi^2$, $R=R(t,r)$ and
$f=f(t,r)$ are arbitrary functions of $t$ and $r$. The metric in
(1) has to satisfy field equations which can be put in a better
form if we put
\be
\dot R^2=f+{F\over R}\Rightarrow \dot R=-\sqrt{f+{F\over R}}
\ee
$F=F(t,r)$, $R=R(t,r)$ and $f(t,r)>-1$ are $C^2$ functions of $t$
and $r$ throughout the cloud. Notation (') and  $(^.)$ are used to
denote partial differentiation with respect to $r$ and $t$.
$F=F(t,r)$ is interpreted as the mass function and for physical
reasons $F(t,r)\ge 0$, $F'(t,r)\ge 0$. $f(t,r)$ may be interpreted
as total energy and classifies the space time as bound, marginally
bound and unbounded as per $f<0$, $f=0$ and $f>0$. For collapse
$\dot R <0$. $R(t,r)$ is called the area radius in the sense that
$4\pi R^2(t,r)$ gives the proper area of the mass shells for a
given value of coordinate $r$ at time $t$ and the area of such a
shell vanishes when $R(t,r)=0$. In this sense curve $t=t(r)$ such
that $R(t(r),r)=0$ describe the singularity in the space-time
where mass shells collapse to zero volume and where curvatures
diverge. To avoid shell crossing we would require $R'>0$. The
specification of arbitrary functions $F$ and $f$ gives a
particular solution of the field equations, for example $F=F(r)$,
$f=f(r)$ gives the TBL models. (For the metric (1) satisfying
field equations in general see Appendix A.)

Our intentions and aim in this paper are certainly not to find
another exact solution of the field equations prescribing a
particular matter field satisfying a particular equation of state,
but to investigate within the framework of general relativity mass
loss when physically allowed reasonable evolution develop into a
singularity naked or covered. The answer to the problem would have
been simple if we had at our disposal both an exact closed
equation of state describing the state of a collapsing matter and
an exact solution of the field equation. However both of these are
little understood in highly dense regions in relativity. Dust
solutions being an exact solution has been widely used for
investigating the occurrence of naked singularities and black
holes(see \cite{D1} and references therein), however as pointed
out earlier it is of limited  use in the context of our work here.
We therefore consider a general physically reasonable matter field
(i.e. type I matter fields) satisfying energy conditions. Let us
consider  matter cloud given by
\be
f(t,r)=-f_0(r)+f_1(t,r),\quad F(t,r)= F_0(r) - f_1(t,r)R \ee Where
$F_0(r)\ge 0, f_0(r)\ge 0$ and $f_1(t,r)\ge 0$ are functions which
are $C^2$ throughout the matter cloud. For type I fields
satisfying energy conditions it follows (For details of the
solution for such a consideration in equation (3), including
energy conditions etc. see Appendix A.)
\be
F_0'\ge f_1'R, 1>f_0-f_1\ge 0,\quad\dot F\le 0,\quad F_0'\ge 0 \ee
Because of equation (3) equation (2) now reads
\be
\dot R=\sqrt{-f_0+{F_0\over R}}
\ee

We now elaborate on equation (3) and such a particular choice.
First note that in case $f_1(t,r)=0$ the space-time given in
metric (1) reduces to TBL dust models. The structure of
differential equation (5) determining $R(t,r)$ remains as in dust
case implying same functional dependence of $R(t,r)$ as a function
of $t$ and $r$ as in dust. The metric in (1)is therefore similar
to TBL model except for the extra term $f_1$ in
$f=f_0(r)-f_1(t,r)$ appearing in the denominator of the $g_{rr}$
term in the metric. The requirement that the space-time be
singularity free,  in the sense that Kritchmann
scalar, principle density and pressures etc. do not diverge prior
to the formation of shell focusing singularity, implies
$f_1=r^nh(t,r),n\ge 2$ with $h(t,r)$ being at least $C^2$
function. Therefore the analysis regarding the formation of a
naked singularity or a black-hole in the gravitational collapse
for such a model remains similar to dust case. The advantage of
such a consideration is that while the analysis and conditions for
occurrence of singularity either naked or covered would be same as
dust, we now have in the space time a type I matter field
satisfying energy conditions with non vanishing pressures and
which could be matched to Vaidya space-time out side the star.

The dust models have been extensively investigated for occurrence
of singularities naked and otherwise. Because of the relevance of
dust models to our work we give, before a general consideration, a
very brief account of dust solutions and use the needed results
while referring the
reader to articles in literature for details \cite{D1}. By
suitably scaling coordinate $r$ at time $t=0$(when the collapse
starts) such that $R(0,r)=r$, we get from equation (5)
\be
t-t_0(r)=-{R^{3/2}\over \sqrt{F_0}}G({R\over r}),
\ee
\be
G(y)={\arcsin{\sqrt{y}}\over y^{3/2}}-{\sqrt{1-y}\over y},
G(0)={2\over 3},\quad t_0(r)= {\pi r^{3/2}\over 2\sqrt{F_0}}
\ee
Where we have used the fact that for the mass shells to start from
rest at $t=0$ $\dot R(0,r)=0\Rightarrow F_0(r)=rf_0(r)$. For
physical reasons we have restricted ourselves to bound cases only
that is $0\ge f>-1$ throughout. In order to avoid shell crossing
we require $t_0'(r)\ge 0$. The function $f_0(r)=f_cr^2g(r),g(0)>0$
is such that $g(r)$ is at least $C^2$ function of $r$ throughout.
This ensures that there are no singularities in the space-time
prior to the formation of first singularity at the center $r=0$
which occurs at time $t=t_s=t_0(0)$ such that $R(t_s,0)=0$. Only
the first central singularity could be naked while all other for
$r>0$ are necessarily covered. The central shell focusing
singularity is globally naked if \cite{D1}
\be
{3\pi f_c\over 4f_{co}^{3/2}}>13+{15\sqrt{3}\over 2}
\ee
Where $f_c>0$ and $f_{co}\ge 0$ are constants given by
\be
f_0(r)=f_cr^2g(r)=f_cr^2(1-f_{co}r^3 + r^4f_{01}(r)) ,
\ee
$f_{01}(r)$ being $C^2$ differentiable  function throughout the
dust cloud. In general $f_0(r)/r^2$ would have terms of the order
of $r$ and $r^2$ as well, however the resulting naked singularity
is gravitationally weak \cite{D1}. For occurrence of strong
curvature singularity $f_0(r)$ must have the form in equation (9).
If the condition in equation (8) is not satisfied than the
singularity is covered (i.e. black hole).

Let us now consider a general scenario. If the  spherically
symmetric star considered above is to radiate energy the most
general external metric is Vaidya space time
\be
ds^2= -(1-{2m(u)\over {\bf r}})du^2 -2dud{\bf r}+ {\bf
r}^2d\Omega^2,
\ee
where the arbitrary function $m(u)$ of the
retarded time $u$ represent the mass of the system out side the
star. The Matching conditions are given in equations (29) to
(33)(see appendix B for details). The motion of the boundary in
Vaidya space-time is given by ${\bf r}=R(t(u),r_b)$ Equation (32)
relates retarded time $u$  to time $t$ . At $t=0, u=u_i$ and as
$t\rightarrow t_{ah}, u\rightarrow \infty$, $t_{ah}$ is the time
when the outermost shell at $r=r_b$ meets the
apparent horizon $R(t_{ah},r_b)=F(t_{ah},r_b)$. Using equation (3) and
(29) we have
\be
F_0(r_b)-f_1(t,r_b)R(t,r_b)=2m(u),
\ee
The matching of the two space-time across the boundary $r=r_b$
restricts the function $f_1(t,r)$ to the form given in equation
(33) and we have for $f_(t,r_b)$
\be
f_1(t,r_b)={4c(r_b)(1-f_0(r_b))\exp({2Y_0(r_b)
\arccos(\sqrt{{R(t,r_b)\over r_b}})}) \over
(1-c(r_b)\exp({2Y_0(r_b)\arccos(\sqrt{{R(t,r_b)\over r_b}})}))^2}
\ee
where functions $c(r), Y_0(r)$ are related to initial pressure
$p_r(r)$ and $p_{\theta}(r)$(see appendix A) via equations (27) and (28)
$f_0(r)$ is an
important quantity in our calculations and we get from
equations (27) and (11)
\be
f_0=r^2p_r+{2M(r)\over r},\quad 2M(r)=k\int{\rho r^2 dr}
\ee
\be
f_0(r_b)={2m_0\over r_b}+ kr_b^2p_r(r_b),\quad m_0=M(r_b)=m(0)
\ee
where the  initial density $\rho\ge 0$ and  pressure $ p_r\ge 0$ and
$p_{\theta}\ge 0$ are $C^2$ functions throughout the cloud. For
physical reasonableness we require that
$\rho'\le 0, p_r'\le 0, p_{\theta}'\le 0$.
Thus we now have a matter cloud with non vanishing
pressures and where the space out side the collapsing star is
described by the Vaidya space time. The mass function $m(u)$ is
related to initial density and pressures of interior matter cloud
making it possible to examine mass loss and that is precisely what
we look for in the next section.

\newpage \section{Mass Loss}

Prior to the onset of collapse the star might appear to be a
normal radiating star to an outside observer receiving radiation
at some rate. These radiation could have been coming for a long
time in the past as they do in most cases. Our interest here,
however, is only to estimate the energy radiated (mass loss)
once the collapse starts.

The exterior metric Out side the moving boundary ${\bf
r}=R(t(u),r_b)$ is given by Vaidya metric. Thus the star with
radius $r=r_b$ at the onset of collapse at $t=0$ continues
radiating till the outermost shell given by $r=r_b$ meets the
surface $R=F$. Actually the surface $R=F$ is characterized as
apparent horizon and represent the boundary beyond which all
causal particles are ingoing. From equation (32)it
follows that as $R(t,r_b)\rightarrow F(t,r_b)$
$u\rightarrow\infty$ thus while the time $t$ is finite for an
inside observer, an out side observer would take infinite
amount of time for the surface of the star to reach the apparent
horizon. 0nce the boundary of the star meets the apparent horizon
there would not be any out going radiation. This is the stage at
which the remaining mass of the collapsing cloud disappears into
the black hole and becomes the mass of the black hole.
The mass of the star at the onset of collapse is given by equation
(14). we have
\be
x_0={2m_0\over r_b}=E_0-T_0,\quad E_0=f_0(r_b),\quad T_0
=kr_b^2p_r(r_b)
\ee
$x_0$ actually is the ratio of the Schwarzschild radius $2m_0$ of
the star with mass $m_o$ at the onset of collapse, to the radius
of the star $r_b$. At the time $t=0$ the boundary of the star
$r=r_b$ starts collapsing. The star continues to radiate till
its boundary $r=r_b$ meets the apparent horizon $R=F$ at time
$t=t_{ah}$ which is given by
\be
R(t_{ah},r_b)=F(t_{ah},r_b)
 \ee
Using equations (3)
\be
1+f_1(t_{ah},r_b)={f_0(r_b)\over R(t_{ah},r_b)}
\ee Using
equations (3), (11) to (16) and eliminating $c(r_b)$ by using
equation (27) we get
\be
{\sqrt{1-x_0}-\sqrt{1-E_0}\over \sqrt{1-x_0}+\sqrt{1-E_0}}={
\sqrt{E_0(1-x_s)}-\sqrt{x_s(1-E_0)}\over \sqrt{E_0(1-x_s)}+\sqrt{x_s(1-E_0)}}
\exp(-2y_0\arccos(\sqrt{x_s}))
\ee
where we have put
\be
x_s={2m_r\over r_b}={2m(\infty)\over r_b},\quad y_0=\sqrt{{1-E_0\over E_0}}
\quad {x_s\over x_o}={m_r\over m_o}
\ee
$2m_r=F(t_{ah},r_b)=2m(\infty)$ is the remnant mass of the star.
The above equation relates final mass
of the collapsing cloud to the initial mass of the star at the
onset of collapse . The value of $E_0$ is the parameter which
determines the exact form of this relation via the above equation.
Thus for a given initial or remnant mass and $E_0$ at the onset of
collapse the remnant mass or the initial mass as the case may be,
can be calculated from equation (18).

It should be noted at this point from equation (8) that condition
for a strong curvature naked singularity or a black hole depend on
the ratio $f_c/f_{co}^{3/2}$, whereas in equation (18)
$E_0=f_0(r_b)$ is the value of the $f_0(r)$(equation (9)) at the
boundary of the star $r=r_b$. Thus for a given value of $E_0$ one
still has the freedom to choose the different set of parameters
$(f_c,f_{co})$ so as to satisfy the condition for either naked
singularity or a black hole. Therefore for any set of initial and
remnant mass in the case of naked singularity resulting in a
certain mass loss one can have a black hole also with a different
set of structure in the form of density and pressure distribution
functions at the onset of collapse. Hence we can conclude that
naked singularity scenario can not radiate more energy than a
black hole scenario. The same is true even if the naked
singularity is weak.

We next consider the case when the total mass, size and the
internal structure of the star is same. Let us consider two stars
with same mass $m_o$, radius $r=r_b$, density distribution
$\rho=\rho_c\rho_o(r)$ and pressure distribution $p_r=p_cp_o(r),
p_{\theta}=p_{\theta c}p_{o \theta} (r)$ at the initial epoch of
collapse. Here only central density $\rho_c$ and central pressure
$p_c$ could be different while the internal structure is kept same
i.e. $\rho_o(r), p_o(r)$ are same. This imply that
for both cases $f_0(r)=f_cr^2g(r)$  where $g(r)$ is kept the same
but different values of $f_c$ are allowed. In fact $f_c= 3\rho_c+
p_c$.
It follows that for $f_c>f_{crit}$ the
collapse would terminate in a black hole while for $f_c\le
f_{crit}$ would end in a naked singularity, $f_{crit}$ being the
critical value determined by equation (8). Physically what it
means that higher central density and pressure at the onset of
collapse lead to black-hole while lesser ones lead the collapse to
naked singularity. Therefore black hole scenarios occur for all
values of $E_0$ greater than certain critical value while less
than the critical value would end in a naked singularity. Equation
(18) implies that for a constant value of $x_0$, $x_s$ decreases
for increasing value of $E_0$( for the typical behavior see Fig
1). Therefore scenarios which end up in black hole would have a
less remnant mass than the naked singularity scenario having the
same initial mass. Hence the collapsing stars which terminate in a
black hole would suffer a higher mass loss and radiate more energy
than the stars which end up in naked singularity having the same
mass, size and same internal structure at the onset of collapse.

Another interesting scenario is where the
collapse scenario is determined completely by a single parameter.
Let us consider at the onset of
collapse, matter cloud satisfying adiabatic equation of state,
i.e. $p_r(r)= p_{\theta}(r) =a\rho(r)$, $a$ being a constant. Here
we consider the two scenarios of naked singularity and black-hole
which occur due to the different values of $a$ and where the total
mass, size, and the density composition of the star are same. We
have
\be
E_0=x_0+a\rho_b
\ee
$\rho_b=\rho(r_b)$ is the initial surface
density of the star. Using equations (9), (27) and (28) the condition for
strong curvature naked singularity given by equation (8 )
becomes.
\be
{3\pi(1+6a)\over 4(1+3a)^{5/2}}\rho_1\ge 13+{15\over 2}\sqrt{3},
\rho_1={\rho'''(0)\over \rho(0)^{{5\over 2}}}
\ee
Thus for a given
$\rho_1$ naked singularity occurs for $a\le a_{crit}$ and a black
hole for $a> a_{crit}$. The behavior of the ratio of remnant mass
$m_r$ to the initial mass $m_o$ remains the same as in previous
consideration. That is for increasing $a$ this ratio decreases
implying that black hole scenario would radiate more. A typical
behavior is shown in Fig 2.

Within the framework of classical general relativity
the overall consideration of the collapsing star( model incorporating
an interior and a exterior space-time) as considered here does not
seem to point towards the significance of the formation of a naked
singularity in respect of energy radiated.
Instead black hole scenario radiate more energy.
Thus the occurrence of a naked singularity may not be physically
more relevant than a
black-hole and even if naked singular scenario does free up a
region of space-time for radiation, the overall energy radiated
does not seem to increase.
We have
considered here a typical model(though the space-time considered
represent quite a wide class). However in general for spherically
symmetric collapse satisfying physically reasonable
conditions as considered here( for example type I field, energy
conditions etc.)the results should be the same qualitatively.

\section{Appendix A.}

In general relativity the metric describing the space time
must satisfy the field equations $G_{ab}=kT_{ab}$,
where $k$, $G_{ab}$ and $T_{ab}$ are the coupling constant,
Einstein and stress energy tensors respectively. For a spherically
symmetric space-time described by the metric in equation (1) we
have
\be
-k_0T_t^t={F'\over R^2R'}-{\dot f\dot R\over R(1+f)}=
{(F_o-f_1R)'\over R^2R'}-{\dot f_1\dot R\over R(1-f_0+f_1)}
\ee
\be
kT_r^r=({\dot F\over R^2\dot R}+{\dot f\over R\dot
 R})={f_1\over R^2},\quad kT_r^t={-\dot f\dot R\over R(1+f)}=
 {\dot -f_1\dot R\over R(1-f_0+f_1)},
\ee
$$2T_{\theta}^{\theta}={(f-R\ddot R-\dot R^2)'\over 2RR'}+{\dot
f\over 1+f}({\dot R\over R}+\ddot f +{\dot R'\over R'}-{3\dot
f\over 2(1+f)}) ={f_1'\over RR'}$$
\be
+{\dot f_1\over
1-f_0+f_1}({\ddot f_1\over 2\dot f_1}- {3\dot f_1\over
2(1-f_0+f_1)}+
{F_0'\over R\dot R R'}-{f_0'\over \dot R R'}-{F_0\over
R^2\dot R})
\ee
Second part of the above give the non vanishing components of the stress
energy tensor for the matter cloud given by equation (3).
We are interested only in type I matter field satisfying energy conditions
\cite{HW,D5} It follows that throughout the cloud
\be
F_0'\ge f_1'R,\dot F\le 0\quad
 {F_0'\over R^2R'}-{f_1'\over RR'}-{2f_1\over R^2}
-{\dot f_1\dot R\over 2R(1-f_0+f_1)} \ge 0
\ee
Using  the notation $\sigma=-T^t_t, P=T^r_r$ and $q=T^t_r$.
\be
\epsilon={\sigma - P +\sqrt{(\sigma+P)^2-4q^2}\over 2},\quad
P_r={-\sigma + P + \sqrt{(\sigma+P)^2-4q^2}\over 2},
P_{\theta},P_{\phi}
\ee
where the four real eigenvalues $\epsilon=\epsilon(t,r),P_r=P_r(t,r)$ and
$P_{\theta}=P_{\phi}=T^{\theta}_{\theta}$ represent the principle
density, principle radial pressure and principle tangential
pressure respectively.
At $t=0$ $T_{tr}=q=0$ and therefore initial density
$\rho(r)=\sigma(0,r)=\epsilon(0,r)$ and initial radial pressure
$p_r(r)=P(0,r)=P_r(0,r)$. Using equations (3), (12),and (22) to (25)
we get after some simplification
\be
kr^2\rho(r)=(rf_0-kr^3p_r)',kr^2p_r=f_1(0,r)\rightarrow
{4c\over(1-c)^2}={kr^2p_r\over 1-f_0}
\ee
\be
p_{\theta}=P_{\theta}(0,r)=
{(r^2p_r)'\over 2r} +{p_rY_0^2f_0\over
2(1-f_0+r^2P_r)}
\ee

\section{Appendix B.}

Nearly all distant objects (stars) in our universe radiate energy
in the form of light like particles. Thus the exterior space time
out side a spherically symmetric star is rightfully described by
the Vaidya space time. The energy for these radiation comes from
within the star. The spherically symmetric matter cloud prior to
the onset of collapse may be radiating energy in the form of
light like particles and in the very distant past may even not be
radiating( i.e. in the very distant past the space time out side
the star may be Schwarzschild). We therefore discuss the related
boundary conditions relevant to our work in this appendix.

We first start with the interior space time metric in equation (1)
for which describe the collapsing spherically symmetric matter cloud
$$ds^2= -dt^2 +{{R'^2}\over{1+f}}dr^2 + R^2d\Omega^2,$$
At the onset of collapse at $t=0$ $R(0,r)=r$. The exterior Vaidya
space time is given by equation (11)
$$ds^2= -(1-{2m(u)\over {\bf r}})du^2 -2dud{\bf r}+ {\bf r}^2d\Omega^2$$
The boundary $\Sigma$
in the internal metric is $r=r_b=constant$. The Darmois conditions
for matching the space-time require that the first and second
fundamental forms of the interior and exterior metric are
continuous across $\Sigma$ \cite{D5}. This is straight forward and
we get the following after using equation (3) and some simplification
\be
{\bf r}=R(t(u),r_b),\quad F(t,r_b)=2m(u)  \ee
\be
{dt\over du}=\sqrt{1+f(t,r_b)}-\sqrt{f(t,r_b)+{F(t,r_b)\over
R(t,r_b)}}
\ee
\be
\dot F|_{r=r_b}=-\left(\dot f R(1-\sqrt{{f+{F\over R}\over
1+f}}\right)_ {r=r_b}
\ee
Using equation (3)
\be
u-u_i=\int_0^t{dt\over
\sqrt{1-f_0(r_b)+f_1(t,r_b)}-\sqrt{f_0(r_b)}\sqrt{{r_b\over
R(t,r_b)}-1}}
\ee
At the onset of collapse at $t=0, u=u_i$.
The apparent horizon is given by $R=F$ and $t=t_{ah}$ is the
time when the boundary of the star $r=r_b$ meets the apparent
horizon, it therefore follows from the above equation (32)
$t\rightarrow t_{ah}, u\rightarrow \infty$ and thus $\infty >u\ge u_i$.
Equation (31) imply that function $f_1(t,r)$ must of the form
\be
f_1(t,r)={4c(r)(1-f_0(r))\exp{(2Y_0(r)
\arccos(\sqrt{{R\over r}}))} \over (1-c(r)\exp{(2Y_0(r)
\arccos(\sqrt{{R\over r}}))})^2} +r^2f_2(t,r), Y_0(r_b)=\sqrt{{1-f_0(r_b)\over f_0
(r_b)}}
\ee
where $f_2(t,r)$ is a arbitrary $C^2$ function through out the cloud such
that $f_2(0,r)=f_2(t,r_b)=0$.
Though not essential from the point of view of our
purpose however we should discuss briefly on the nature of the
space time prior to gravitational collapse not only from the
theoretical interest but also to complete the discussion.
This intermediate state prior to the collapse should
preferably be a compact star in equilibrium with out side space
time preferably a radiating metric like Vaidya space times(or some
generalization of it)matched at some fix boundary $r=r_b$. The exact
form of interior metric for $t<0$ would depend actually on the
specific form of the arbitrary function $f_1(t,r)$ and its
behavior near $t=0$. However for the sake of completeness we
consider the same. Let us consider the interior metric of the star
for $t<0$ which may be described by
\be
ds^2= -AdT^2 +Bdr^2 + r^2d\Omega^2,
\ee
where $A(t,r)>0$ $B(t,r)>0$ are at least $C^2$ functions through
out the cloud $r_b\ge r\ge 0$. Taking into account the fact that
at $t=0$ $R(0,r)=r$, $\dot R(t,r)=0$ for the metric in equation
(1) Darmois matching conditions for these interior metric at the
surface $t=0$ are \cite{D5}
\be
B(0,r)=f_0(r)-f_1(0,r),\quad {\dot B(0,r)\over
\sqrt{A(0,r)}}={-\dot f_1(0,r)\over 1-f_0+r^2p_r}
\ee
Prior to collapse the exterior metric outside the static boundary $r=r_b=
constant$ of the star could be Vaidya space-time.
\be
ds^2= (1-{2M_p(u)\over{\it r}})du^2 -2du d{\it r}+  {\it r}^2d\Omega^2,
\ee
Matching conditions at the boundary $r=r_b$ require
\be
{dT\over du}=\sqrt{{1-{2M_p(u)\over r_b}\over A(T,r_b)}}
\ee
\be
{1\over B(T,r_b)}=1-{2M_(u)\over r_b}
\ee
\be
A(T,r_b)A'(T,r_b)=(1-{2M_p(u)\over r_b}){2M_p(u)\over
r_b^2}-{2\over r_b}{dM_p(u)\over du}
\ee
Thus a set of functions
$(A,B)$ satisfying the above conditions at the onset of collapse
would describe the interior and exterior of the star prior the
onset of collapse. Because of the generality of the two
arbitrary functions $A(t,r)$ and
$B(t,r)$ the suitable solution satisfying boundary conditions
would exist for a given $f_1(t,r)$. Though the exact
expressions for $A(T,r)$ and $B(T,r)$ would depend on the exact
form of $f_1$ we can consider a simple example.
Consider For the
function $f_1$ given in equation (33) such that $f_2=0$
The space-time prior to onset of collapse therefore is given by
\be
A=(1+a_1(r)M_p(1-{2M_p\over r_b}-r_b{dM_p\over du})^{1/2}
\ee
\be
B= (1-B_0(1-{B_1\over b_1})-{2B_1\over b_1r_b}{dM_p\over
du})^{-1}
\ee
where $a_1(r)$ and $M_p(u)$ are arbitrary functions of $r$ and $u$
respectively such that
\be
a_1'(r_b)=2r_b,\quad M_p(u_i)=M_0, ({dM_p\over
du})_{u=u_i}=-(E_0-x_0)\sqrt{1-x_0}
\ee
\be
B_0=B_0(r)=f_0(r)-r^2p_r(r),\quad, b_0=B_0(r_b)
\ee
\be
B_1=B_1(r)=-r^2p_r\sqrt{1-f_0+r^2p_r},\quad b_1=B_1(r_b)
\ee

\newpage
\begin{figure}[h]
\epsfbox{fig1.epsi}
\caption{ Plot showing the ratio of the remnant mass to the
initial mass(i.e. ${m_r\over m_o}$) vs $E_0$. The critical value
being $E_0=0.202$ for a star of radius $r_b=10m_o\rightarrow x0=0.2$
with initial mass $m_o$ . Thus region $E_0>0.202$ correspond to
black-hole region while $E_0,\le 0.202$ is the naked singular
region.}
\end{figure}

\newpage
\begin{figure}[]
\epsfbox{fig2.epsi}
\caption{Graph showing the mass ratio $m_r/m_o$ vs $a$. Here
$a\approx 0.001$ is the critical value of $a$. Thus for a star of
radius $r_b\approx 45m_0\rightarrow x0\approx 0.04408$ and with
surface density $\rho(r_b)\approx 0.001$. Region $a> 0.001$ is
black-hole region while the other is naked singular region. }
\end{figure}


\begin{thebibliography}{99}
\bibitem{D4} In fact all known and observed physical matter
field are of type I except the directed radiations( Vaidya
space-time) which are type II. A type I matter field is
characterized by the existence of four real eigenvalues
$\lambda_{(i)}$ of the energy momentum tensor i.e.
$T^{ab}=\Sigma_{i=0}^{i=3} \lambda_iE^a_{(i)}E^b_{(i)}$,
$E^a_{(i)}$ being an orthonormal basis. $\lambda_0=\rho$
associated with timelike basis vector $E^{(0)}$ represent the
principle density while $P_1=\lambda_1,P_2=\lambda_2, P_3=
\lambda_3$ are principal presseurs. The energy conditions require
$\rho\ge 0,\rho+P_1\ge 0$, $\rho+P_2\ge 0$ and $\rho+P_3\ge 0$.
For more details see \cite{HW}.
\end{thebibliography}
\end{document}